\documentstyle[epsfig,12pt]{article}
\begin{document}
\begin{center}
{\bf{ $SU(2)$ Yang-Mills Theory in Savvidy Background at Finite Temperature
and Chemical Potential}}

\vspace{1.5cm}

R.Parthasarathy{\footnote{e-mail:sarathy@imsc.res.in}} 
and Alok Kumar{\footnote{e-mail:alok@imsc.res.in}}
 
\vspace{0.5cm}

The Institute of Mathematical Sciences \\
C.P.T.Road, Taramani Post \\
Chennai 600113, India. \\
\end{center} 

\vspace{1.5cm}

{\noindent{\it{Abstract}}}

\vspace{0.5cm}

The one-loop effective energy density of a pure $SU(2)$ Yang-Mills
theory in the Savvidy background, at finite temperature and chemical
potential is examined with emphasis on the unstable modes. 
After identifying the stable and unstable modes,
the stable modes are treated in the quadratic approximation. For the
unstable modes, the full expansion including the cubic and the quartic
terms in the fluctuations is used. The functional integrals for the
unstable modes are evaluated and added to the results for the stable
modes. The resulting energy density is found to be {\it{real}},
coinciding with the real part of the energy density in the quadratic
approximation of earlier study. There is now {\it{no imaginary part.}}
Numerical results are presented for the variation of the energy density
with temperature for various choices of the chemical potential. 

\vspace{1.0cm}


\newpage

{\noindent{\bf{I.INTRODUCTION}}}

\vspace{0.5cm}

The ground state of $SU(2)$ Yang-Mills theory in a covariantly constant
chromomagnetic field in the third color direction ($F_{12}^3=H$) had
been studied by Savvidy [1]. The important result obtained was the
one-loop effective energy density was lower than the perturbative vacuum
and the theory has a gluon condensate, $<|F^a_{\mu\nu}F^{\mu\nu a}|>\neq
0$. However, Nielson and Olesen [2] pointed out that the one-loop
effective energy density had an imaginary part due to the lowest Landau
level. Various studies were devoted to this aspect [3]. In recent
times, gauge invariance arguments to exclude the imaginary part [4] and
a stable vacuum by dynamically generating mass term for the off-diagonal
gluons [5] have been proposed. In all these studies, only the terms quadratic
in the fluctuations are kept, throwing out the cubic and quartic terms
in the fluctuations. As early as 1983, the Savvidy vacuum within the
framework of the background field method was examined by Flory [6] and
later by Kay [7] treating the unstable modes including the cubic and
quartic terms and the possibility of obtaining real effective energy
density was pointed out. The present authors with Kay [8] have
explicitly shown at zero temperature, the resulting energy density,
when the cubic and the quartic terms in the fluctuations are included
for the unstable modes, has {\it{no imaginary part}} and coincided with
the real part of the earlier calculations. Briefly, the stable modes are
treated in the quadratic approximation and for the unstable modes, the
functional integral is evaluated keeping the cubic and quartic terms
along with the quadratic term. This result (Eqn.30 of [8]) has no
imaginary part and when added to the result of the stable modes, gives
the effective energy density which is real. The calculations leading to
this result are gauge invariant [8]. 

\vspace{0.5cm}

The Savvidy vacuum at finite temperature has been studied by many authors
[9 - 15]. In all these studies, the quadratic approximation has been
used. As a result, the effective energy density involved an imaginary
part, dependent on the temperature. This is serious as it persists at
high temperatures. It has been observed by Meisinger and Ogilvie [15]
that with the introduction of a non-trivial Polyakov loop, specified by
$\phi$, it is possible to stabilize the vacuum, if $\beta \sqrt{gH}<
\phi <2\pi - \beta \sqrt{gH}$, where $\beta=1/{kT}$, $H$ is the
chromomagnetic background in the third color direction, as for this
range of $\phi$, the imaginary part becomes zero. However, the imaginary
part is non-zero at the global minimum. So, in the understanding of the
finite temperature behaviour, the imaginary part severely inhibits the
progress.

\vspace{0.5cm}

It is the purpose of this work to extend our method [8] at zero
temperature, to finite temperature with chemical potential, by
separating the unstable modes and including the cubic and quartic terms
in these modes in the evaluation of the functional integral and adding
to the contribution of the stable modes in the quadratic approximation.
The motivation is to get real effective energy density. Another purpose
is to resolve a discrepancy in the analytical expressions for the
one-loop energy density between [14] and [15]. The discrepancy is an
interchange of $J_1$ and $Y_1$ and also a relative sign between the two
$K_1$ functions in the energy density. Our results show that the finite
part of the effective energy density is real. There is no imaginary
part. The real energy density coincides with the real part of [15] which
therefore resolves the discrepancy alluded above in favour of [15]. 

\vspace{0.5cm}

The next section develops the formalism for the evaluation of the
effective energy density. The stable and the unstable modes are
separately treated. Section.III provides the details of handling the
unstable modes including the cubic and quartic terms in the expansion.
The results are added to the contribution of the stable modes and the
full expression for the effective energy density is exhibited. Section
IV contains the details of the numerical evaluation of the effective
enerdy density for various temperatures. 

\vspace{1.0cm}

{\noindent{\bf{II. EFFECTIVE ENERGY DENSITY IN THE BACKGROUND FIELD METHOD
(BFM)}}}

\vspace{0.5cm}

The Euclidean functional integral for an $SU(2)$ pure YM theory is
\begin{eqnarray}
Z&=& \int [dA_{\mu}^a] e^S,
\end{eqnarray}
where 
\begin{eqnarray}
S&=&\int d^4x \{-\frac{1}{4} F_{\mu\nu}^aF^{\mu\nu a}\},
\end{eqnarray}
and 
\begin{eqnarray}
F^a_{\mu\nu}&=&{\partial}_{\mu}A^a_{\nu}-{\partial}_{\nu}A^a_{\mu}+g
{\epsilon}^{abc}A^b_{\mu}A^c_{\nu}.
\end{eqnarray}
Expanding $A^a_{\mu}={\bar{A}}^a_{\mu}+a^a_{\mu}$ with
${\bar{A}}^a_{\mu}$ as the classical background field satisfying the
equation of motion ${\bar{D}}^{ab}_{\mu}{\bar{F}}^b_{\mu\nu}=0$, with
${\bar{D}}^{ab}_{\mu}={\partial}_{\mu}{\delta}^{ab}+g{\epsilon}^{acb}
{\bar{A}}^c_{\mu}$ as the {\it{background covariant derivative}},
${\bar{F}}^a_{\mu\nu}$ is same as (3) with ${\bar{A}}^a_{\mu}$ and using
the background gauge 
\begin{eqnarray}
{\bar{D}}^{ab}_{\mu}a^b_{\mu}&=&0,
\end{eqnarray}
we have
\begin{eqnarray}
Z&=& \int [da^a_{\mu}] e^{S'},
\end{eqnarray}  
with 
\begin{eqnarray}
S'&=&\int d^4x \Big(
-\frac{1}{4}{\bar{F}}^a_{\mu\nu}{\bar{F}}^a_{\mu\nu}+\frac{1}{2}
a^a_{\mu}{\Theta}^{ac}_{\mu\nu}a^c_{\nu}+g{\epsilon}^{acd}({\bar{D}}
^{ae}_{\nu}a^e_{\mu})a^c_{\mu}a^d_{\nu} \nonumber \\
&-& \frac{g^2}{4}\{ (a^a_{\mu}a^a_{\mu})^2-a^a_{\mu}a^c_{\mu}a^a_{\nu}
a^c_{\nu}\}-\ell og \ det(-{\bar{D}}^{ab}_{\mu}{\bar{D}}^{bc}_{\mu}
)\Big),
\end{eqnarray}    
where 
\begin{eqnarray}
{\Theta}^{ac}_{\mu\nu}&=&({\bar{D}}^{ab}_{\lambda}{\bar{D}}^{bc}_
{\lambda}){\delta}_{\mu\nu}+2g {\epsilon}^{aec}{\bar{F}}^e_{\mu\nu}.
\end{eqnarray}

In arriving at (5) and (6), we have introduced the gauge fixing and the
ghost lagrangian for the background gauge (4) and integrated the ghost
fields, resulting in the last term in (6). The expansion in (6) is
{\it{exact}}. If the cubic and the quadric terms in $a^a_{\mu}$ are
neglected in (6), then the effective potential will be given by 
\begin{eqnarray}       
{\Gamma}^{1-loop}&=&\frac{1}{2} Tr\ \ell og\ det(-{\Theta}^{ac}_{\mu\nu}
)-Tr\ \ell og\ det(-{\bar{D}}^{ab}_{\mu}{\bar{D}}^{bc}_{\mu}). 
\end{eqnarray}

\vspace{0.5cm}

The Savvidy background is chromomagnetic in the third color direction
and is given by 
\begin{eqnarray}
{\bar{A}}^a_0=0 &;& {\bar{A}}^a_i\ =\ {\delta}^{a3}(-\frac{Hy}{2},
\frac{Hx}{2},0),
\end{eqnarray}
which gives ${\bar{F}}^3_{12}=H$ and solves the classical equation of
motion, ${\bar{D}}^{ab}_{\mu}{\bar{F}}^b_{\mu\nu}=0$. We introduce the
chemical potential ($\mu$) as a background field, following Loewe,
 Mendizabel and Rojas [16], namely
\begin{eqnarray}
{\bar{B}}^a_{\mu}&=&\frac{\mu}{g}v_{\mu}{\delta}^{a3}, \ \ \ \ v_{\mu}
=(1,0,0,0).
\end{eqnarray}
This method, introduced in [16], has the advantage that as chemical
potentials are not introduced as Lagrange multipliers in BFM, it is not
necessary to compute the conserved charges. In [16], only the background
(10) was considered. In our case we combine (9) and (10) as
\begin{eqnarray}
{\bar{\cal{A}}}^a_{\mu}&=&{\delta}^{a3}(\frac{\mu}{g},-\frac{Hy}{2},
\frac{Hx}{2},0),
\end{eqnarray}
which gives again ${\bar{F}}^3_{12}=H$ as the non-vanishing background
field strength and {\it{which solves the classical equation of motion.}}
In this method of introducing the chemical potential (10,11), it will
play the role as the Polyakov loop specified by a constant
${\bar{\cal{A}}}^a_4$ field in the third color direction as in [15] with
the identification of $\phi=\beta \mu$. As in [8] (eqn.10), we notice
that (7) gives 
\begin{eqnarray}
{\Theta}^{ac}_{44}={\Theta}^{ac}_{33}&=&{\bar{D}}^{ab}_{\lambda}
{\bar{D}}^{bc}_{\lambda},
\end{eqnarray}
where now
\begin{eqnarray}
{\bar{D}}^{ab}_{\lambda}&=&{\partial}_{\lambda}{\delta}^{ab}+g{\epsilon}
^{a3b}{\bar{A}}^3_{\lambda}+\mu {\epsilon}^{a3b}v_{\lambda},
\end{eqnarray}
where ${\bar{A}}^3_{\lambda}$ is given by (9) and $v_{\lambda}$ in (10).
The relation (12) gives the important result that the contributions of
${\Theta}^{ac}_{44}$ and ${\Theta}^{ac}_{33}$ to ${\Gamma}^{1-loop}$ in
(8), cancel the ghost contribution in (8). Using (11) in (7), it is easy
to see that
\begin{eqnarray}
{\Theta}^{ac}_{41}={\Theta}^{ac}_{42}=&
{\Theta}^{ac}_{43}&={\Theta}^{ac}_{31}={\Theta}^{ac}_{32}=0.
\end{eqnarray}     

\vspace{0.5cm}

The remaining eigenvalues are from ${\Theta}^{ac}_{ij}$ for $i,j=1,2$
only. The eigenmodes and the corresponding eigenvalues are: 
\begin{eqnarray}
(a^1_1+ia^2_1)+i(a^1_2+ia^2_2)&:&(k_4+\mu)^2+k_3^2+(2N+1)gH-2gH,
\nonumber \\
(a^1_1+ia^2_1)-i(a^1_2+ia^2_2)&:&(k_4+\mu)^2+k_3^2+(2N+1)gH+2gH,
\nonumber \\
(a^1_1-ia^2_1)+i(a^1_2-ia^2_2)&:&(k_4-\mu)^2+k_3^2+(2N+1)gH+2gH,
\nonumber \\
(a^1_1-ia^2_1)-i(a^1_2-ia^2_2)&:&(k_4-\mu)^2+k_3^2+(2N+1)gH-2gH,
\nonumber \\ 
 a^3_i &:& k_4^2+k_3^2+k_2^2+k_1^2; \ i=1,2,
\end{eqnarray}
where $N=0,1,2, \cdots$ is the harmonic oscillator (in the $x-y$ plane)
quantum number.

\vspace{0.5cm}

We now pass on to the finite temperature case by replacing [17] $k_4$ by
$\frac{2\pi n}{\beta}$ and $\int_{-\infty}^{\infty} dk_4$ by
$\sum_{n=-\infty}^{\infty}$. With this, the effective potential in the
"quadratic approximation" becomes
\begin{eqnarray}
{\Gamma}^{1-loop}&=&2\times \frac{1}{2}\times
\frac{1}{\beta}\sum_{n=-\infty}^{\infty}\int \frac{d^3k}{(2\pi)^3}
\ell og\Big(\Big(\frac{4{\pi}^2n^2}{{\beta}^2}+{\vec{k}}^2
\Big)/{\Lambda}^2\Big)
\nonumber \\
&+&\frac{1}{2}\times \frac{1}{\beta}\Big(\frac{gH}{2\pi}\Big)\sum_
{n=-\infty}^{\infty}\ \sum_{N=0}^{\infty}\int_{-\infty}^{\infty}
\frac{dk_3}{2\pi} \nonumber \\
& &\Big( \ell og\frac{1}{{\Lambda}^2}\{ (\frac{2\pi
n}{\beta}+\mu)^2+k_3^2+(2N+1)gH-2gH\} \nonumber \\
&+&\ell og\frac{1}{{\Lambda}^2}\{ (\frac{2\pi n}{\beta}+\mu)^2+k_3^2+
(2N+1)gH+2gH\} \nonumber \\
&+&\ell og\frac{1}{{\Lambda}^2}\{ (\frac{2\pi n}{\beta}-\mu)^2+k_3^2+
(2N+1)gH+2gH\} \nonumber \\
&+&\ell og\frac{1}{{\Lambda}^2}\{ (\frac{2\pi n}{\beta}-\mu)^2+k_3^2+
(2N+1)gH-2gH\}\Big),     
\end{eqnarray}      
where the first term corresponds to the last eigenvalue in (15) and the
remaining correspond respectively to the first four eigenvalues in (15).
The pre-factor $\frac{gH}{2\pi}$ is the harmonic oscillator degeneracy
and the pre-factor 2 in the first term accounts for the two modes in the
last term in (15) for $i=1,2$. $\Lambda$ is the dimensionful parameter
to render the argument of the logarithms dimensionless. We suppress this
hereafter.

\vspace{0.5cm}

This expression (16) agrees with Eqn.8 of [15] with their $\phi$
replaced by $\mu\beta$ and with Eqn.(2.16) of Ninomiya and Sakai [12]
with $\mu=0$. It can be seen from (15), that for $k_3=0$ and $n=0;N=0$,
the eigenvalues first and the fourth become ${\mu}^2-gH$ and therefore
to avoid negative eigenvalues $\mu > \sqrt{gH}$. However, for $n=1$, the
fourth eigenvalue becomes $(\frac{2\pi}{\beta}-\mu)^2-gH$ (with N=0 and
$k_3=0$) and if this is to remain positive, then $\mu <
\frac{2\pi}{\beta}-\sqrt{gH}$. With this, the first eigenvalue remains
positive. So for $\sqrt{gH}< \mu < \frac{2\pi}{\beta}-\sqrt{gH}$, it is
possible to avoid the negative eigenvalues and hence the instability.
However, the instability enters at the global minimum [15]. We will
return to this in the next section. Now, we proceed to evaluate (16). In
order to exhibit the contribution from the unstable modes, we consider
the {\it{second logarithm}} (without its prefactors) in (16) and write that 
as [8]
\begin{eqnarray}
L_2&=&-\sum_{n=-\infty}^{\infty}\sum_{N=0}^{\infty}\int_{-\infty}
^{\infty}\frac{dk_3}{2\pi}\ \int_{0}^{\infty}dt\ t^{-1}\ e^{-t\{
 (\frac{2\pi n}{\beta}+\mu)^2+k_3^2+A\}},
\end{eqnarray}
where $A=2NgH-gH$. After performing the $k_3$ integration and the sum
over $n$, we find
\begin{eqnarray}
L_2&=&-\frac{1}{2\sqrt{\pi}}\sum_{N=0}^{\infty}\int_{0}^{\infty}\ dt 
\ t^{-\frac{3}{2}}\ {\theta}_3\Big( \frac{2\mu t i}{\beta},\frac{4\pi i t}
{{\beta}^2}\Big)\ e^{-t({\mu}^2+A)}, \nonumber 
\end{eqnarray}
where 
\begin{eqnarray}
{\theta}_3(z,\tau)&=&\sum_{n=-\infty}^{\infty} e^{i\pi\tau n^2} \ 
e^{2\pi i n z}. \nonumber 
\end{eqnarray}
Now, using the property of the ${\theta}_3$-function [18],
\begin{eqnarray}
{\theta}_3(z,i\tau)&=& {\tau}^{-\frac{1}{2}}\ e^{-\pi z^2/{\tau}}
\ {\theta}_3(\frac{z}{i\tau},\frac{i}{\tau}), \nonumber 
\end{eqnarray}
we get
\begin{eqnarray}
L_2&=&-\frac{\beta}{4\pi}\sum_{N=0}^{\infty}\int_{0}^{\infty}\ dt\ t^{-2}
\ {\theta}_3(\frac{\mu \beta}{2\pi}, \frac{i{\beta}^2}{4\pi t})
\ e^{-tA}. \nonumber 
\end{eqnarray}
Including the prefactor in (16), the contribution to ${\Gamma}^{1-loop}$
from $L_2$ is 
\begin{eqnarray}
{\Gamma}^{1-loop}(L_2)&=&-\frac{gH}{16{\pi}^2}\ \sum_{N=0}^{\infty} \int
_{0}^{\infty}\ dt\ t^{-2}\ {\theta}_3(\frac{\mu\beta}{2\pi},\frac{
i{\beta}^2}{4\pi t})\ e^{-tA}. \nonumber 
\end{eqnarray}
The unstable mode corresponds to $N=0$. Splitting the above sum over $N$
as for $N=0$ and $N=1,2,\cdots ,\infty$, we find
\begin{eqnarray}
{\Gamma}^{1-loop}(L_2)&=&-\frac{gH}{16{\pi}^2}\int_{0}^{\infty}\ 
dt\ t^{-2}
\ {\theta}_3(\frac{\mu\beta}{2\pi},\frac{i{\beta}^2}{4\pi t})\Big(
e^{tgH}+\frac{e^{-tgH}}{1-e^{-2tgH}}\Big),
\end{eqnarray}
where the first term in $\Big(\cdots \Big)$ is the contribution from the
unstable mode.

\vspace{0.5cm}

The third term in (16) has no negative eigenvalue and writing this as in
(17), the sum over $N=0,1,2,\cdots ,\infty$ is performed to give
\begin{eqnarray}
{\Gamma}^{1-loop}(L_3)&=&-\frac{gH}{16{\pi}^2}\int_{0}^{\infty}\ dt \ 
t^{-2}\ {\theta}_3(\frac{\mu\beta}{2\pi},\frac{i{\beta}^2}{4\pi t})\ 
\frac{e^{-3tgH}}{1-e^{-2tgH}}. 
\end{eqnarray}
The fourth term in (16) is the same as the third term except for $\mu$
replaced by $-\mu$. So also the fifth term and the second term are same
but for this replacement. As ${\theta}_3(z,\tau)={\theta}_3(-z,\tau)$
[18], we have 
\begin{eqnarray}
\sum_{j=2}^{5}{\Gamma}^{1-loop}(L_j)&=&-\frac{gH}{8{\pi}^2}\int_{0}
^{\infty}dt\ t^{-2}\ {\theta}_3\Big(\frac{\mu\beta}{2\pi},
\frac{i{\beta}^2}
{4\pi t}\Big)\Big( e^{tgH}+\frac{e^{-tgH}+e^{-3tgH}}{1-e^{-2tgH}}\Big),
\nonumber \\
&=&-\frac{gH}{8{\pi}^2}\int_0^{\infty}dt\
t^{-2}{\theta}_3\Big(\frac{\mu\beta}{2\pi},\frac{i{\beta}^2}{4\pi t}\Big)
\Big(
e^{tgH}-e^{-tgH}+\frac{2e^{-tgH}}{1-e^{-2tgH}}\Big), \nonumber \\
& &  
\end{eqnarray}    

\vspace{0.5cm}

Now ${\theta}_3\Big( \frac{\mu\beta}{2\pi},\frac{i{\beta}^2}{4\pi
t}\Big)$ can be rewritten using
\begin{eqnarray} 
{\theta}_3(z,\tau)&=&\sum_{\ell=-\infty}^{\infty}\ e^{i\pi\tau {\ell}^2}
\ e^{2\pi i z\ell}, \nonumber 
\end{eqnarray}
as
\begin{eqnarray}
{\theta}_3\Big( \frac{\mu\beta}{2\pi},\frac{i{\beta}^2}{4\pi t}\Big) &=&
1+2\sum_{\ell=1}^{\infty} cos(\mu\beta\ell)\
e^{-\frac{{\beta}^2{\ell}^2}{4t}}. \nonumber 
\end{eqnarray}
The exponentials in (20) can be written as
$\frac{cosh(2tgH)}{sinh(tgH)}$ and so (20) becomes
\begin{eqnarray}
&-\frac{gH}{8{\pi}^2}\int_{0}^{\infty}dt\ t^{-2}\
\frac{cosh(2tgH)}{sinh(tgH)}\Big(1+2\sum_{\ell=1}^{\infty}
cos(\mu\beta\ell)\ e^{-\frac{{\beta}^2{\ell}^2}{4t}}\Big).&
\end{eqnarray}
  
The first term in (16) is the familiar term and its finite part is 
$-\frac{{\pi}^2}{45{\beta}^4}$. Thus the one-loop potential is
\begin{eqnarray}
{\Gamma}^{1-loop}&=&-\frac{{\pi}^2}{45{\beta}^4} \nonumber \\
&-&\frac{(gH)^2}{8{\pi}^2}\int_{0}^{\infty} d\tau \ {\tau}^{-2}\ 
\frac{cosh(2\tau)}{sinh\tau}\Big(1+2\sum_{\ell=1}^{\infty}
cos(\mu\beta\ell)e^{-\frac{{\beta}^2{\ell}^2gH}{4\tau}}\Big),
\end{eqnarray}
where $\tau=gHt$. The zero temperature part is reproduced. In [8], this
was evaluated treating the unstable modes carefully including the cubic
and the quartic terms and the effective energy density has no imaginary
part. It was found to be,
\begin{eqnarray}
{\cal{E}}_{1-loop}(T=0)&=&\frac{H^2}{2}+\frac{11(gH)^2}{48{\pi}^2}\Big(
\ell og\Big(\frac{gH}{{\Lambda}^2}\Big)-\frac{1}{2}\Big).
\end{eqnarray}

\vspace{0.5cm}

We now proceed to evaluate the finite temperature part of (22),
\begin{eqnarray}
{\Gamma}^{1-loop}(T)&=& -\frac{{\pi}^2}{45{\beta}^4} \nonumber \\
 &-&\frac{(gH)^2}{4{\pi}^2}\int_0^{\infty}d\tau\ {\tau}^{-2}\
\frac{cosh{2\tau}}{sinh\tau}\sum_{\ell=1}^{\infty}cos(\mu\beta\ell)
\ e^{-\frac{{\beta}^2{\ell}^2gH}{4\tau}}, \nonumber \\
&=&-\frac{{\pi}^2}{45{\beta}^4} \nonumber \\
&-&\frac{(gH)^2}{4{\pi}^2}\int_{0}^{\infty}d\tau\ {\tau}^{-2}\Big( 
e^{\tau}-e^{-\tau}+\frac{2e^{-\tau}}{1-2e^{-2\tau}}\Big) 
\sum_{\ell=1}^{\infty}cos(\mu\beta\ell)e^{-\frac{{\beta}^2{\ell}^2gH}{4
\tau}}, \nonumber \\
& &  
\end{eqnarray}
where we exhibit the {\it{unstable mode contribution}} explicitly by the
first term in $\Big(\cdots \Big)$. Expanding
\begin{eqnarray}
\frac{1}{1-e^{-2\tau}}&=&\sum_{n=0}^{\infty}\ e^{-2n\tau}\ =\ e^{-2\tau}
\ +\ \sum_{n=1}^{\infty}e^{-2n\tau}, \nonumber 
\end{eqnarray}
and introducing
\begin{eqnarray}
I_1&=&\int_{0}^{\infty}\ d\tau\ {\tau}^{-2}\ e^{\tau}\
e^{-\frac{{\beta}^2{\ell}^2gH}{4\tau}},
\end{eqnarray}
\begin{eqnarray}
I_2&=&\int_{0}^{\infty}\ d\tau\ {\tau}^{-2}\ e^{-\tau}\
e^{-\frac{{\beta}^2{\ell}^2gH}{4\tau}}, 
\end{eqnarray}
\begin{eqnarray}
I_3&=&2\int_{0}^{\infty}\ d\tau\ {\tau}^{-2}\ e^{-(2n+1)\tau}\
e^{-\frac{{\beta}^2{\ell}^2gH}{4\tau}},
\end{eqnarray}
the finite temperature part of (24) is written as
\begin{eqnarray}
{\Gamma}^{1-loop}&=&-\frac{{\pi}^2}{45{\beta}^4} \nonumber \\
&-&\frac{(gH)^2}{4{\pi}^2}\sum_{\ell=1}^{\infty}cos(\mu\beta\ell)\
 \Big( I_1+I_2+I_3\Big),
\end{eqnarray}
in which $I_1$ is from the {\it{$N=0$ unstable mode}}. The integrals in
(27) are to be evaluated.

\vspace{0.5cm}

In $I_1$ (25), we perform a Wick rotation to arrive at 
\begin{eqnarray}
I_1&=&i^{-1} \int_0^{\infty}\ dt\ t^{-2}\
e^{i\Big(t+\frac{{\beta}^2{\ell}^2gH}{4t}\Big)}, \nonumber \\
&=&\frac{2\pi i}{\beta \ell\sqrt{gH}}\ H_1^{(1)}\Big(\beta\ell\sqrt{gH}
\Big),
\end{eqnarray}
where $H_1^{(1)}$ is the Hankel function of the first kind [19]. 

$I_2$ and $I_3$ integrals are evaluated using [19] 
\begin{eqnarray}
K_{\nu}(xz)&=&\frac{z^{\nu}}{2}\int_0^{\infty}\ t^{-\nu-1}\
e^{-\frac{x}{2}(t+\frac{z^2}{t})}\ dt, 
\end{eqnarray}
as 
\begin{eqnarray}
I_2&=&\frac{4}{\beta\ell\sqrt{gH}}\ K_1\Big(\beta\ell\sqrt{gH}\Big),
\end{eqnarray}
\begin{eqnarray}
I_3&=&8\sum_{n=1}^{\infty}\ \frac{\sqrt{(2n+1)}}{\beta\ell\sqrt{gH}}\
K_1\Big(\sqrt{(2n+1)}\beta\ell\sqrt{gH}\Big),
\end{eqnarray}                            
so that the finite part of (28) becomes
\begin{eqnarray}
{\Gamma}^{1-loop}(T)&=&\frac{{\pi}^2}{45{\beta}^4} \nonumber \\
&-&\frac{(gH)^{\frac{3}{2}}}{\beta{\pi}^2}\sum_{\ell=0}^{\infty}
\frac{cos(\mu\beta\ell)}{\ell} \nonumber \\
&\times &\Big(\frac{i\pi}{2}H_1^{(1)}\Big(\beta\ell\sqrt{gH}\Big)+K_1\Big(
\beta\ell\sqrt{gH}\Big) \nonumber \\ 
&+&2\sum_{n=1}^{\infty}\sqrt{(2n+1)}
K_1\Big(\sqrt{2n+1}\beta\ell\sqrt{gH}\Big)\Big).
\end{eqnarray}

\vspace{0.5cm}

Using $H_1^{(1)}(\beta\ell\sqrt{gH})\ =\ J_1(\beta\ell\sqrt{gH}) + 
iY_1(\beta\ell\sqrt{gH})$, we find that our result (33) agrees with
Meisinger and Ogilvie [15] who used a different method to evaluate (16).
The interchange of $J_1$ and $Y_1$ and a relative sign between the two
$K_1$ functions in Starinets, Vshivtsev, and Zhukovskii [14] are
incorrect. From (28) and (33), it is seen that the unstable mode ($N=0$)
contributions are contained in $I_1$ and explicitly
\begin{eqnarray}
{\Gamma}^{1-loop}_{unstable}(T)&=&-\frac{(gH)^{\frac{3}{2}}}{\beta
{\pi}^2}\sum_{\ell=1}^{\infty}\frac{cos(\mu\beta\ell)}{\ell}\Big(
-\frac{\pi}{2}Y_1(\beta\ell\sqrt{gH})+\frac{i\pi}{2}J_1(\beta\ell
\sqrt{gH})\Big). \nonumber \\
& &  
\end{eqnarray}
The imaginary part above is remniscent of the zero temperature
situation. In the later case, we [8] have treated the unstable modes
including the cubic and quartic terms in (6) and showed then the
contribution is real. In view of the important difficulties raising from
the imaginary part at {\it{finite temperature}}, we, in the next
section, treat the unstable modes at finite temperature including the
cubic and the quartic terms in the expansion (6).

\vspace{1.0cm}

{\noindent{\bf{III. FINITE TEMPERATURE UNSTABLE MODES-INCLUSION OF CUBIC AND
QUARTIC TERMS}}}

\vspace{0.5cm}

There are two unstable modes when the harmonic oscillator quantum number
$N=0$, as seen from (15). They are the first and the fourth line in (15)
with $N=0$. As the expression (16) was obtained in the "quadratic
approximation", we cannot use (16) for the unstable modes when we want
to include the cubic and the quartic terms in the unstable modes. So, we
take the equation (5) for $Z$ but confine ourselves to the unstable
modes only here. Without loss of generality we will take $\mu=0$, zero
chemical potential. The normalized unstable mode eigenfunctions are
\begin{eqnarray}
{\phi}_{k_3,n}(x)&=&\sqrt{\frac{gH}{2\pi}}e^{-\frac{gH}{4}(x^2_1+x^2_2)}
\ \frac{1}{\sqrt{L_3\beta}}\ e^{-i(k_3x_3+\frac{2\pi n}{\beta}x_4)},
\end{eqnarray}
where the first exponential is the ground statet wavefunction ($N=0$) of
the harmonic oscillator in the $(x_1-x_2)$ plane, the second exponential
is the box-normalized plane wave in the $x_3$ direction and the
$S^1$-harmonics in the $x_4$-direction. The index $n$ is the Matsubaro
index and is not the same $n$ in (33) as the later index originated in
the expansion of $\frac{1}{1-e^{-2\tau}}$ in the expression above (25).
The unstable eigenmodes in (15) are then $c(k_3,n){\phi}_{k_3,n}(x)$ and 
$c(k_3,n)^*{\phi}^*_{k_3,n}(x)$ respectively.

\vspace{0.5cm}

The unstable modes involve Lorentz indices 1 and 2 (as can be seen in
(15)) and the $SU(2)$ indices 1 and 2, since the classical background in
(11) is in the third isospin direction. Consequently, the cubic term
${\epsilon}^{acd}({\bar{D}}^{ae}_{\nu}a^e_{\mu})\ a^c_{\mu}a^d_{\nu}$
vanishes as in [8]. The terms quartic in $a^a_{\mu}$ in (5) are
simplified to $\frac{1}{8}\ a^+_+\ a^-_-\ a^+_+\ a^-_-$. Then, the full
partition function for the unstable modes, from (5) and (6) is
\begin{eqnarray}
Z_{unstable}&=&\int[da^a_{\mu}]\ e^{-\int
d^4x\{a_u(k_3^2+k_4^2-gH)a_u+\frac{g^2}{32}a_u^4\}}, 
\end{eqnarray}      
where $a_u$ stands for $a^+_+$ and $a^-_-$. From (36), the unstable
modes for $k_3^2+(\frac{2\pi n}{\beta})^2\ <\ gH$ render the quadratic
term in $a_u$ in the exponent of (36) divergent. {\it{However, the
quartic term in $a_u$ in the exponent of (36) provides the necessary and
the crucial convergence.}} Thus the overall integral over $a_u$ will be
convergent. Now expanding $a_u$ in terms of the eigenfunctions (35) and
carrying out the $d^4x$ integration in the exponent, the quadratic term
becomes 
\begin{eqnarray}
&\{ k_3^2+(\frac{2\pi n}{\beta})^2-gH\} c^2(k_3,n), & 
\end{eqnarray}    
and the quartic term becomes
\begin{eqnarray}  
& \frac{g^2}{32}\ \frac{(gH)^2}{4{\pi}^2}\ \frac{\pi}{gH}c^4(k_3,n)\ =\
 \frac{g^3H}{128\pi}c^4(k_3,n),&  
\end{eqnarray}
where we have taken all the four ${\phi}_{k_3,n}(x)$ having the same
$k_3,\ n$. Then, we obtain
\begin{eqnarray}
Z_{unstable}&=&\Big(\int {\prod}_{k_3,n}\ dc(k_3,n)\ e^{\{(gH-k_3^2-
(\frac{2\pi n}{\beta})^2)c^2-\frac{g^2(gH)}{128\pi}c^4\}}\Big)^D,
\end{eqnarray}
where $D$ is the degeneracy factor $D=\frac{gH}{2{\pi}^2}V$ with $V$ as
the spatial volume. Now, introducing $\hat{c}=\sqrt{gH}c$ and
${\hat{k}}_3=\frac{k_3}{\sqrt{gH}}$, (39) is written as
\begin{eqnarray}
Z_{unstable}&=&{\Big( \frac{1}{\sqrt{gH}}\int {\prod}_{k_3,n}\
d\hat{c}\ e^{\{(1-{\hat{k}}_3^2-(\frac{2\pi
n}{\beta\sqrt{gH}})^2){\hat{c}}^2-\frac{g^2}{128\pi (gH)}{\hat{c}}^4
\}}\Big)}^D. 
\end{eqnarray}

\vspace{0.5cm}

The contribution of the complete unstable modes to the energy density

$(
-\frac{1}{\beta V}\ell og {Z_{unstable}})$ 
is then
\begin{eqnarray}
&-\frac{gH\sqrt{gH}}{2{\pi}^2\beta}\int dk_3\sum_n \ \ell og J(k_3,n),
\end{eqnarray}
where 
\begin{eqnarray}
J(k_3,n)&=&\int_{-\infty}^{\infty}d\hat{c}\ e^{\{(1-{\hat{k}}_3^2-
(\frac{2\pi n}{\beta\sqrt{gH}})^2){\hat{c}}^2-\frac{g^2}{128\pi(gH)}
{\hat{c}}^4\}}.
\end{eqnarray}
In (42), if the quartic term is neglected, then (41) will give (34). We
retain the crucial quartic term contribution. The integral over
$d\hat{c}$ is convergent {\it{irrespective}} of the sign of the
coefficient of the ${\hat{c}}^2$ term. It is evaluated using [19]
\begin{eqnarray}
\int_0^{\infty}e^{-({\beta}^2x^4+2{\gamma}^2x^2)}dx&=&2^{-\frac{3}{2}}
\Big(\frac{\gamma}{\beta}\Big)\ e^{\frac{{\gamma}^4}{2{\beta}^2}}\ 
K_{\frac{1}{4}}\Big(\frac{{\gamma}^4}{2{\beta}^2}\Big), \nonumber 
\end{eqnarray}
with 
\begin{eqnarray}
{\beta}^2&=&\frac{g^2}{128\pi(gH)}, \nonumber \\
{\gamma}^2&=&\frac{1}{2}\Big( {\hat{k}}_3^2+(\frac{2\pi n}{\beta\sqrt
{gH}})^2-1\Big), \nonumber 
\end{eqnarray}
Then (42) becomes
\begin{eqnarray}
J({\hat{k}}_3,n)&=& \frac{8\sqrt{\pi gH}}{\sqrt{2}g}\Big( {\hat{k}}_3^2+
(\frac{2\pi n}{\beta\sqrt{gH}})^2-1\Big)^{\frac{1}{2}}\ e^{\frac{16\pi
(gH)}{g^2}\{{\hat{k}}_3^2+(\frac{2\pi n}{\beta\sqrt{gH}})^2-1\}^2}
\nonumber \\
&\times & K_{\frac{1}{4}}\Big(\frac{16\pi(gH)}{g^2}\{{\hat{k}}_3^2+
(\frac{2\pi n}{\beta\sqrt{gH}})^2-1\}^2\Big).
\end{eqnarray}   

\vspace{0.5cm}

When ${\hat{k}}_3^2+(\frac{2\pi n}{\beta\sqrt{gH}})^2<1$ (that is where
the instability arose), the above expression using [20] (since $n$
starts from $1$, the argument of $K_{\frac{1}{4}}$ will be small)  
\begin{eqnarray}
K_{\nu}(x)&\rightarrow & \frac{2^{\nu-1}\Gamma(\nu)}{x^{\nu}};\ \ \ small\ x,
\nonumber 
\end{eqnarray}
becomes
\begin{eqnarray} 
J({\hat{k}}_3,n)&\simeq &\frac{8\sqrt{\pi gH}}{\sqrt{2}g}\Big(
{\hat{k}}_3^2 +(\frac{2\pi n}{\beta\sqrt{gH}})^2-1\Big)^{\frac{1}{2}} 
\nonumber \\
&\times & \frac{2^{-\frac{3}{4}}\Gamma(\frac{1}{4})}{\{{\hat{k}}_3^2
+(\frac{2\pi n}{\beta\sqrt{gH}})^2-1\}^{\frac{1}{2}}}\ \frac{\sqrt{g}}
{2(\pi gH)^{\frac{1}{4}}}, \nonumber \\
&=& \frac{2\sqrt{2}}{\sqrt{g}}(\pi gH)^{\frac{1}{4}}2^{-\frac{3}{4}}
\Gamma(\frac{1}{4}).
\end{eqnarray}
In particular the radical ${\{{\hat{k}}_3^2+(\frac{2\pi
n}{\beta\sqrt{gH}})^2-1\}}^{\frac{1}{2}}$ {\it{gets cancelled.}} {\it{This
result is real. The imaginary part coming from the radical gets
cancelled by the contribution from $K_{\frac{1}{4}}$. This is made
possible by the inclusion of the quartic term.}} When (44) is used in (41), the
${\hat{k}}_3$-integration and the sum over $n$ (constrained by 
${\hat{k}}_3^2+(\frac{2\pi n}{\beta\sqrt{gH}})^2<1$) produces an
uninteresting term which is omitted.

\vspace{0.5cm}

When ${\hat{k}}_3^2+(\frac{2\pi n}{\beta\sqrt{gH}})^2>1$, the expression
(43) takes the form, using [20],
\begin{eqnarray}
K_{\nu}(x)&\rightarrow & \sqrt{\frac{\pi}{2x}}\ e^{-x};\ \ \ large\ x,  
\nonumber   
\end{eqnarray}
as
\begin{eqnarray}
J({\hat{k}}_3,n)&=&{\Big( {\hat{k}}_3^2+(\frac{2\pi
n}{\beta\sqrt{gH}})^2-1\Big)}^{-\frac{1}{2}}.
\end{eqnarray}
Then (41) is evaluated as:
\begin{eqnarray}
\int d{\hat{k}}_3\sum_n\ell og
J({\hat{k}}_3,n)&=&-\frac{1}{2}\int_{1}^{\infty}\sum_n\ell og\{
{\hat{k}}_3^2+(\frac{2\pi n}{\beta\sqrt{gH}})^2-1\}, \nonumber \\
&=&-\frac{1}{2}\int_{1}^{\infty}d{\hat{k}}_3\ell og \{cosh(\beta
\sqrt{({\hat{k}}_3^2-1)gH})-1\}, \nonumber \\
&=&-\frac{1}{2}\int_{1}^{\infty}d{\hat{k}}_3\Big(\beta\sqrt{({\hat{
k}}_3^2-1)gH}\ +\ 2\ell og(1-e^{-\beta\sqrt{({\hat{k}}_3^2-1)gH}})\Big),
\nonumber \\
&=&-\frac{1}{2}\beta\int_{1}^{\infty}d{\hat{k}}_3\sqrt{({\hat{k}}_3^2-1)
gH}\ -\ \frac{\pi}{2}\sum_{n=1}^{\infty}\frac{1}{n}Y_1(n\beta\sqrt{gH})
\ +\ I_1, \nonumber \\
& &  
\end{eqnarray}
where
\begin{eqnarray}
I_1&=&-\sqrt{gH}\sum_{n=1}^{\infty}\int_0^{\frac{\pi}{2}}
cos(n\beta\sqrt{gH}cos\theta)cos{\theta} \ d\theta.
\end{eqnarray}

\vspace{0.5cm}

The first term in (46), when used in (41), produces a
$\beta$-independent contribution and neglected for finite temperature
effects. The integral $I_1$ in (47) is evaluated [19] to be
\begin{eqnarray}
I_1&=&-\sqrt{gH}\sum_{n=1}^{\infty}\Big(-\{-1+\sum_{k=1}^{\infty}
\frac{(-1)^{k-1}(n\beta\sqrt{gH})^{2k}}{1\cdot (2^2-1)\cdot (4^2-1)
\cdots ((2k)^2-1)}\}\Big), \nonumber 
\end{eqnarray}
which is not contributing to the finite part of the energy density.
Thus, the unstable mode contributions to the finite $\beta$-dependent
part of the energy density is found to be
\begin{eqnarray}
&\frac{gH\sqrt{gH}}{2{\pi}^2\beta}\cdot \frac{\pi}{2}\sum_{n=1}^{\infty}
\frac{1}{n}\ Y_1(n\beta\sqrt{gH}),
\end{eqnarray}
which is just the real part of (34). There is no imaginary part. This is
due to the inclusion of the cubic and the quartic terms in the unstable
modes. Thus, the difficulties associated with the imaginary part are not
due to the intrinsic property of the $SU(2)$ chromomagnetic ground state
but due to the use of the gaussian approximation. This reaffirms our
earlier study [8] of the same system at zero temperature.  

\vspace{0.5cm}

The complete expression of the energy density of the $SU(2)$
chromomagnetic state, including the zero contribution from [8] is
\begin{eqnarray}
{\cal{E}}&=&\frac{H^2}{2}+\frac{11(gH)^2}{48{\pi}^2}\{\ell og\Big(
\frac{gH}{{\Lambda}^2}\Big)-\frac{1}{2}\} \nonumber \\
&+&\frac{{\pi}^2}{45{\beta}^4} \nonumber \\
&+&\frac{(gH)^{\frac{3}{2}}}{\beta{\pi}^2}\sum_{\ell=1}^{\infty}
\frac{cos(\mu\beta\ell)}{\ell}\Big(-\frac{\pi}{2}Y_1(\beta\ell\sqrt{gH})
\nonumber \\
&+&K_1(\beta\ell\sqrt{gH})+2\sum_{n=1}^{\infty}\sqrt{2n+1}\ 
K_1(\sqrt{2n+1}\beta\ell\sqrt{gH})\Big). 
\end{eqnarray}  
The finite temperature part agrees with (the real part of) [15]. A
similar result was found for the zero temperature case in [8].         
    
\vspace{1.0cm}

{\noindent{\bf{IV. NUMERICAL RESULTS AND DISCUSSION}}}

\vspace{0.5cm}

The expression for the effective energy density (49) involves summation
over $\ell$ and in most studies, the high and low temperature behaviours
were examined. Even then, the summation is very involved as can be seen
from [14] and [15]. As the high and low temperature expressions are
given in [14] and [15] (also in 12]), we attempt to evaluate (49)
numerically. First, it can be seen from [21], the $K_1$ functions have a
fall-off to zero when the argument is greater than 5. On the other hand
the $Y_1$ function is oscillatory with decreasing amplitude. Second, we
used the "polynomial approximations" for these functions given in [21]
and verified that they are good by computing these functions for various
values of $x$ from 0.05 to large values and comparing them with the
tables of these functions. Then, we used MATLAB to find the values of
$Y_1(x)$ and $K_1(x)$ for various values of $x$ which agree with the
previous method. Third, we need to evaluate the sums in (49).   

For the $Y_1$ function appearing in (49), we
found the typical sum $\sum_{\ell=1}^{\infty}\ \frac{Y_1(\ell x)}{\ell}$
converged to a steady value for $\ell x$ upto 200 whereas for the sum
$\sum_{\ell=1}^{\infty} \frac{K_1(\ell x)}{\ell}$, a steady value is
reached when $\ell x=20$. These allow us to choose ${\ell}_{max}=200/x$
for the $Y_1$ sum while ${\ell}_{max}=20/x$ for the sum involving $K_1$.
Keeping ${\ell}_{max}$ as 20/x, in the last $K_1$ sum, the $n$ sum was
carried out from $n=1$ to $n_{max}$ with
$n_{max}=\frac{1}{2}\{\frac{{\ell}_{max}^2}{{\ell}^2}-1\}$. In order to
evaluate the temperature variation of (49), we first set
$\beta=\frac{a}{\sqrt{gH}}$ and $\mu=b\sqrt{gH}$. Then the temperature
dependent part of (49) becomes
\begin{eqnarray}
\frac{{\cal{E}}_T}{(gH)^2}&=&\frac{{\pi}^2}{45a^4}+\frac{1}{{\pi}^2a}
\sum_{\ell =1}^{\infty}\ \frac{cos(ab\ell)}{\ell}\Big(-\frac{\pi}{2}
Y_1(a\ell) \nonumber \\
&+&K_1(a\ell)+2\sum_{n=1}^{\infty}\sqrt{2n+1}K_1(a\ell \sqrt{2n+1})
\Big).
\end{eqnarray}

\vspace{0.5cm}

In Fig.1, we have plotted $\frac{{\cal{E}}_T}{(gH)^2}$ with $T\ =\
\frac{\sqrt{gH}}{k}\ \frac{1}{a}$, that is, $T$ in units of
$\frac{\sqrt{gH}}{k}$, for $b=0,1,2,3$. For $b=0$, zero chemical
potential, the variation is smooth apart from small oscillatory
behaviour at low temperatures. For $b=2,3$, the variation shows a
minimum and then raising smoothly. At high temperatures, the behaviour
is like that of non-interacting relativistic gas. In [15], the Polyakov
loop is measured in terms of $\phi$ which in our notation is $\mu\beta$
and that is $ab$. In the sense that $a$ is varied, it is not possible to
relate directly our results to [15]. However, the importance of the
chemical potential is seen in Fig.1. A non-zero chemical potential or
non-zero $\phi$ triggers a possible deconfinement phase transition. Our
variation is qualitatively in agreement with [15] for their "real part".

\vspace{0.5cm}

Now, we wish to examine the inclusion of the cubic and quartic terms for
all the modes. It can be seen from (15), the stable eigenvalues being
distinctly different from the unstable eigenvalues for a given $N$. So,
it is justifiable to consider the corresponding eigenmodes as
orthogonal. Then, from (15), it follows that the cubic terms will vanish
for the stable modes as well. The resulting full expression can be
evaluated as in (43) with explicit $N$ appearing. When the logarithm is
taken, as in (41), the finite part will remain unaltered. The situation
for the unstable modes is different in the sense the troublesome
imaginary part does not appear in (44).

\vspace{0.5cm}

To summarize, we have considered the one-loop effective energy density
of a pure $SU(2)$ Yang-Mills theory in the Savvidy background, at finite
temperature and chemical potential. The unstable modes are treated by
keeping the cubic and the quartic terms in the fluctuations. This result
is added to the contribution from the stable modes. There is no
imaginary part. The variation of the energy density for a given
chromomagnetic background with temperature is studied numerically. When
the chemical potential is non-zero, the variation shows a minimum which
is (roughly) interpretted as indicating a deconfinement phase
transition. At high temperatures, the behaviour is like that of a
relativistic gas.                          

\vspace{1.0cm}

{\noindent{\bf{Acknowledgements}}}

\vspace{0.5cm}

The authors thank Dr.G.S.Moni for helping the numerical computations.
One of us (A.K) acknowledges with thanks the award of the Junior
Research Fellowship by IMSc. 

\vspace{1.0cm}

{\noindent{\bf{References}}}

\vspace{0.5cm}

\begin{enumerate}
\item G.K.Savvidy, Phys.Lett. {\bf{B71}} (1977) 133.
\item N.K.Nielsen and P.Olesen, Nucl.Phys. {\bf{B144}} (1978) 376.
\item H.B.Nielsen and M.Ninomiya, Nucl.Phys. {\bf{B156}} (1979) 1; \\
      H.B.Nielsen and P.Olesen, Nucl.Phys. {\bf{B160}} (1979) 380; \\
      J.Ambjorn and P.Olesen, Nucl.Phys. {\bf{B70}} (1980) 60;265; \\
      R.Anishetty, J.Phys.G:Nucl.Phys. {\bf{10}} (1984) 423;439. 
\item Y.M.Cho, {\it{Gauge Invariance and stability of SNO vacuum in
QCD}}, hep-th/0409247.
\item K.-I.Kondo, {\it{Magnetic condensation, Abelian dominance and
instability of Savvidy vacuum in Yang-Mills theory}}, hep-th/0410024.
\item C.A.Flory, {\it{Covariant constant chromomagnetic fields and
elimination of the one-loop instabilities}}, SLAC-PUB-3744, October
1983.        
\item D.Kay, {\it{Unstable modes, zero modes, and phase transition in
QCD}}, Ph.D Thesis, Simon Fraser University, August 1985.
\item D.Kay, A.Kumar and R.Parthasarathy, Mod.Phys.Lett. {\bf{A20}}
(2005) 1655.
\item J.J.Kapusta, Nucl.Phys. {\bf{B190 (FS3)}} (1981) 425.
\item B.Muller and J.Rafelski, Phys.Lett. {\bf{B101}} (1981) 111.
\item J.Chakrabarti, Phys.Rev. {\bf{D24}} (1981) 2232; \\
      M.Reuter and W.Dittrich, Phys.Lett. {\bf{B144}} (1984) 99.
\item M.Ninomiya and N.Sakai, Nucl.Phys. {\bf{B190 (FS3)}} (1981) 316.
\item A.Cabo, O.K.Kalashnikov and A.E.Shabad, Nucl.Phys. {\bf{B185}}
(1981) 473.
\item A.O.Starinets, A.S.Vshivtsev and V.C.Zhukovsky, Phys.Lett.
{\bf{B322}} (1994) 403.
\item P.N.Meisinger and M.C.Ogilvie, Phys.Rev. {\bf{D66}} (2002) 105006.
\item M.Loewe, S.Mendizabel and J.C.Rojas, Phys.Lett. {\bf{B635}} (2006)
213.
\item J.J.Kapusta, {\it{Finite temperature field theory}}, Cambridge
University Press, 1989.
\item Bateman Manuscript, Vol.II, McGraw-Hill, 1953.
\item I.S.Gradshteyn and I.M.Ryzhik, {\it{Table of Integrals, Series and
Products}}, Academic Press, 1965.
\item N.N.Lebedev, {\it{Special Functions and their Applications}},
Prentice-Hall, 1965.
\item M.Abramowitz and I.A.Stegun, {\it{Handbook of Mathematical
Functions}}, Dover Publications, Inc., New York, 1968.
\end{enumerate}        
\begin{figure}
\begin{center}{\hbox{\epsfig{figure=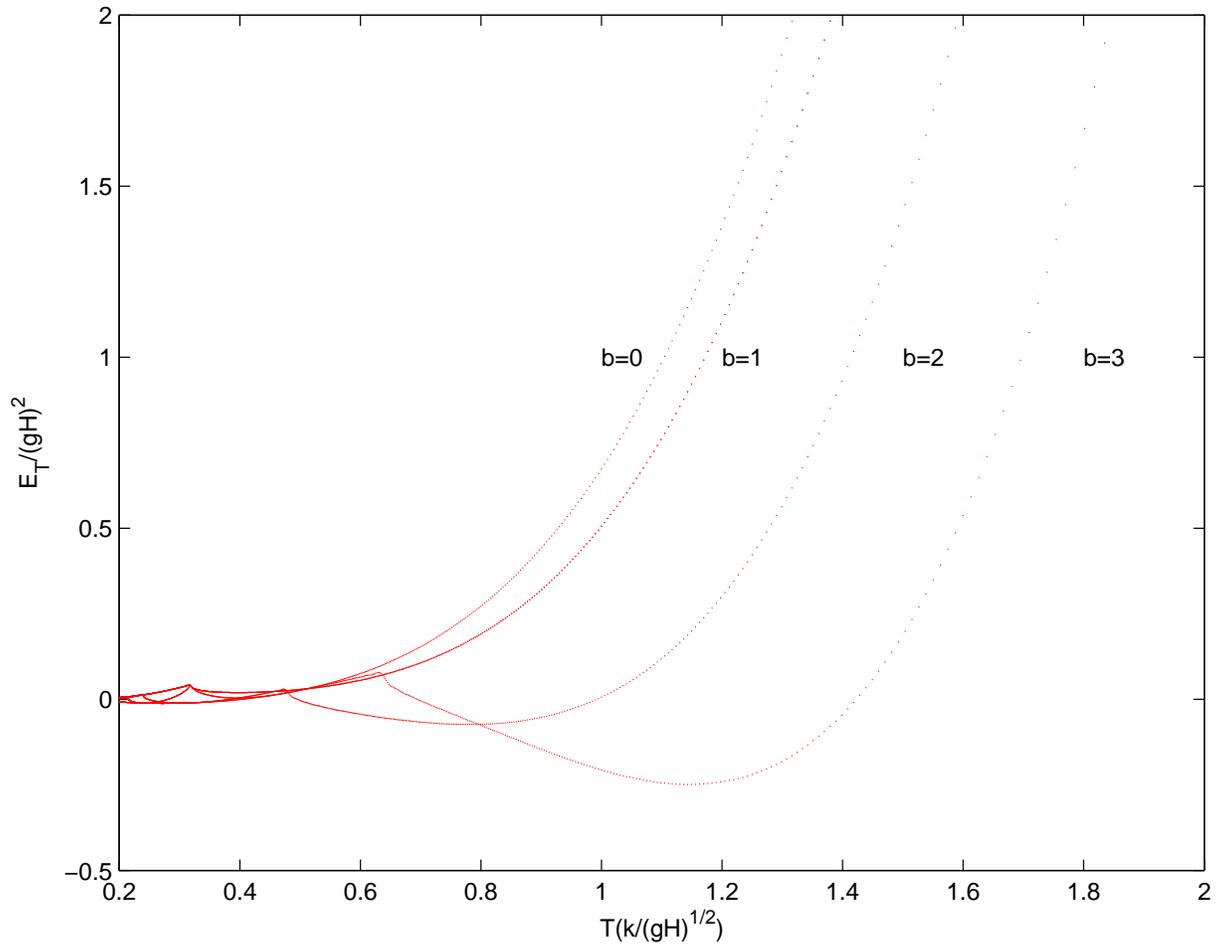,height=5in}}} 
\end{center}
\caption{Variation of scaled energy density with scaled temperature}
\end{figure}                            
\end{document}